\documentclass[iop]{emulateapj}
\usepackage{graphicx}
\usepackage{epsfig}
\usepackage{epstopdf}
\usepackage[colorlinks=true,urlcolor=blue,citecolor=blue,linkcolor=blue]{hyperref}
\usepackage[T1]{fontenc}
\usepackage{amsmath}

\begin{document}
\author{J. de la Cruz Rodr\'iguez\altaffilmark{1}}
\title{Heating of the magnetic chromosphere: observational constraints from \ion{Ca}{2}~$\lambda8542$ spectra}
\author{B. De Pontieu\altaffilmark{2}}
\author{M. Carlsson\altaffilmark{3}}
\author{L.H.M. Rouppe van der Voort\altaffilmark{3}}

\affil{\altaffilmark{1}Department of Physics and Astronomy, Uppsala University, Box 516, SE-75120 Uppsala, Sweden}

\affil{\altaffilmark{2}Lockheed Martin Solar \& Astrophysics Lab, Org.\ A021S,
  Bldg.\ 252, 3251 Hanover Street Palo Alto, CA~94304 USA}

\affil{\altaffilmark{3}Institute of Theoretical Astrophysics,
  University of Oslo,   %
  P.O. Box 1029 Blindern, N-0315 Oslo, Norway}

\begin{abstract}
  The heating of the Sun's chromosphere remains poorly understood. 
  While progress has been made on understanding what drives the quiet
  Sun internetwork chromosphere, chromospheric heating in strong
  magnetic field regions continues to present a difficult challenge, mostly because of a lack
  of observational constraints. We use high-resolution 
  spectropolarimetric data from the Swedish 1-m Solar Telescope to
  identify the location and spatio-temporal properties
  of heating in the magnetic chromosphere. In particular, we report the existence of raised-core
  spectral line profiles in the \ion{Ca}{2}~$\lambda8542$ line. These
  profiles are characterized by the absence of an absorption line
  core,  showing a quasi-flat profile between $\lambda\approx\pm
  0.5$~\AA, and are abundant close to magnetic bright-points and
  plage. Comparison with 3D MHD simulations indicates that such
  profiles occur when the line-of-sight goes through an
  "elevated temperature canopy"
  associated with the expansion with height of the
  magnetic field of flux concentrations. This temperature canopy 
  in the simulations is caused by ohmic dissipation where there are strong
  magnetic field gradients. The raised-core profiles are thus indicators of locations of increased
  chromospheric heating. We characterize the location and temporal and spatial
  properties of such profiles in our observations, thus providing
  much stricter constraints on theoretical models of chromospheric heating
  mechanisms than before.
\end{abstract}

\section{Introduction} 
\label{sec:intro}
The heating of the Sun's chromosphere requires more
  than an order of magnitude more mechanical energy flux than the
  corona and heliosphere combined \citep{Anderson1989,Withbroe1977}.
  Nevertheless, the heating mechanism
  powering the chromosphere remains elusive, especially in the
  magnetic chromosphere in and around network and plage regions \citep[e.g.,][]{2010judge}. This
  is in part because most studies have focused on quiet-Sun
  internetwork regions
  \citep[e.g.,][]{Carlsson1997,Straus2008}.
  In addition, most
  theoretical studies have avoided detailed comparisons with
  chromospheric diagnostics, which are most often formed under non-LTE
  and non-equilibrium
  conditions and thus difficult to model, instead focusing on simply comparing the energy flux
  associated with specific physical heating mechanisms
  \citep[e.g.][]{Goodman2010} to the
  canonical values derived from semi-empirical 1D hydrostatic models of
  the chromosphere \citep{Vernazza1981,Fontenla1993}. Such a limited
  comparison does not properly capture the chromospheric conditions,
  which are significantly different from the 1D VAL and FAL models \cite[in
  terms of dynamics and structuring, see, e.g.,][]{Carlsson1997}. 

\begin{figure*}[t]
\includegraphics[width=\hsize]{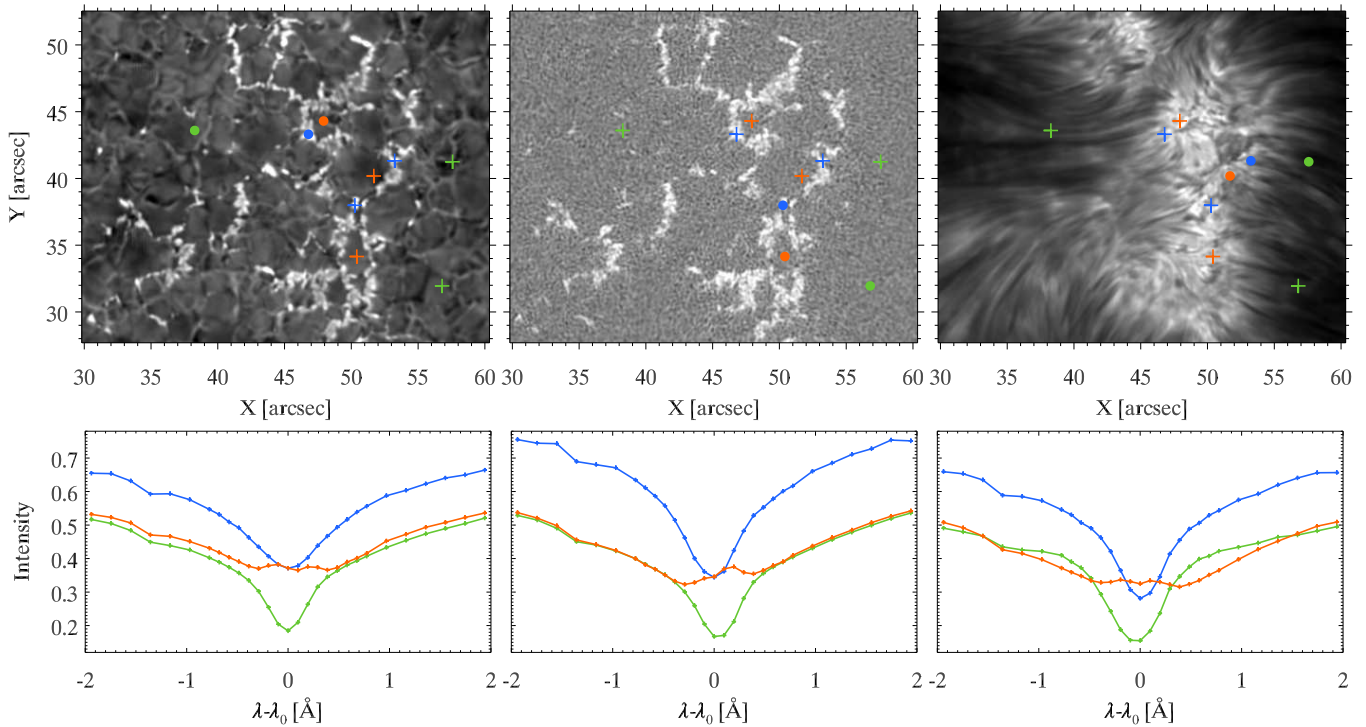}
\caption{\ion{Ca}{2}~$\lambda8542$ filtergrams ({\it top-row}): Stokes~$I$ at
  $-1055$~m\AA\ from line center ({\it left}), Stokes~$V$
at the same wavelength ({\it middle}) and the line core intensity ({\it right}).
The color markers indicate the location
of hand-picked quiet-Sun profiles ({\it green}), RC profiles ({\it orange})
and bright-point profiles ({\it blue}). The circular markers indicate the
location of the intensity profiles shown in each of the panels below (same color coding).}
\label{fig:f1}
\end{figure*}

  Here we report on new observational constraints on chromospheric
  heating derived from the \ion{Ca}{2}~$\lambda8542$ spectral line.
During the past decade, the \ion{Ca}{2} infrared (IR) triplet lines ($\lambda8498$,
$\lambda8542$, $\lambda8662$) have become popular chromospheric
diagnostics in solar studies. In particular, the \ion{Ca}{2}~$\lambda8542$ line has been used in a variety of
different chromospheric studies, for example, to determine the disk
counterpart of spicules \citep{Rouppe2009}, analyze the energy flux
carried by acoustic waves into the lower atmosphere \citep{2009vecchio}, study 
torsional motions in the atmosphere \citep{BDP2012,2012wedemeyer}, or to carry out spectropolarimetric studies in
sunspots \citep{2000socas-navarro-sci, 2001lopez-ariste,2012kleint}.

\citet{2009cauzzi} suggested a connection
between chromospheric heating and the width of the \ion{H}{1}~$\lambda6563$
(H$\alpha$) line, { which is found to be strongly sensitive to
changes in temperature. 
Furthermore, their maps of the line core intensity of \ion{Ca}{2}~$\lambda8542$
show strong similarities with their H$\alpha$ line-width maps. }
Further support for a relationship between core brightness of Ca
lines and elevated temperatures is provided by \citet{2012henriques},
who used the intensity of \ion{Ca}{2}~$\lambda3968$
as a measurement of the upper-photospheric temperature.

Although the velocity field and temperature stratification in the
solar atmosphere produce an imprint in the spatially-resolved profiles
from the \ion{Ca}{2}~IR lines, they usually remain absorption
lines with a conspicuous line core. Interestingly,
\citet{2007pietarilaa} report some examples of reversals in the
NLTE core of the \ion{Ca}{2}~$\lambda8542$ line, which are later associated
with elevated chromospheric temperatures in \citet{2007pietarilab}. However the
mechanism producing this increase in temperature or the
spatio-temporal properties are not
investigated. These raised-core (RC hereafter) profiles are found in
pixels with \emph{intermediate} magnetic flux in their observations.

In this letter, we use full-Stokes 
 \ion{Ca}{2}~$\lambda8542$
data to show the ubiquity of such RC profiles in the vicinity of
strong magnetic fields. We report on the
peculiar spatial distribution of the RC profiles, and analyze their spatial and
temporal properties to provide constraints on chromospheric heating models.
Finally, we propose a scenario that explains the profile shapes, using
synthetic spectral line profiles from a 3D radiative MHD simulation.


\section{Observations}
\label{sec:observations}
The observations are of a bipolar active region
(AR10998) at approximately $205\arcsec$ 
heliocentric distance. The
datasets were acquired at the Swedish 1-m Solar Telescope (SST)  
using the CRisp Imaging SpectroPolarimeter
\citep[CRISP,][]{2008ApJ...689L..69S} on 14 Jun 2008 at 08:36:54.

Our data consists of a 41 minute time-series including 225 full-Stokes line-scans (time steps) of
the \ion{Ca}{2}~$\lambda8542$ line, with a temporal cadence of 11
seconds. The line was sampled at 29 line positions across the range
$\pm 1.94$~\AA \ from the line center, with a spectral sampling of
97~m\AA\
in the core up to $\pm0.78$~\AA, and 194~m\AA\ in the outer wings. 

%


{ The data are processed using the image
restoration technique Multi-Object-Multi-Frame-Blind-
Deconvolution \citep[MOMFBD,][]{2005SoPh..228..191V}, which 
allows to achieve a spatial resolution of $\lambda/D=$0\farcs18.
Instrumental polarization is compensated using the 
2D calibration described in \citet{2011schnerr} and the SST
model at 854~nm from \citet{2010delacruz}.

We have made extensive use of CRISPEX \citep{2012vissers}
a tool for the exploration of CRISP datasets.}

\subsection{The raised-core profiles}\label{sec:rcp}

\begin{figure*}
\includegraphics[width=\hsize]{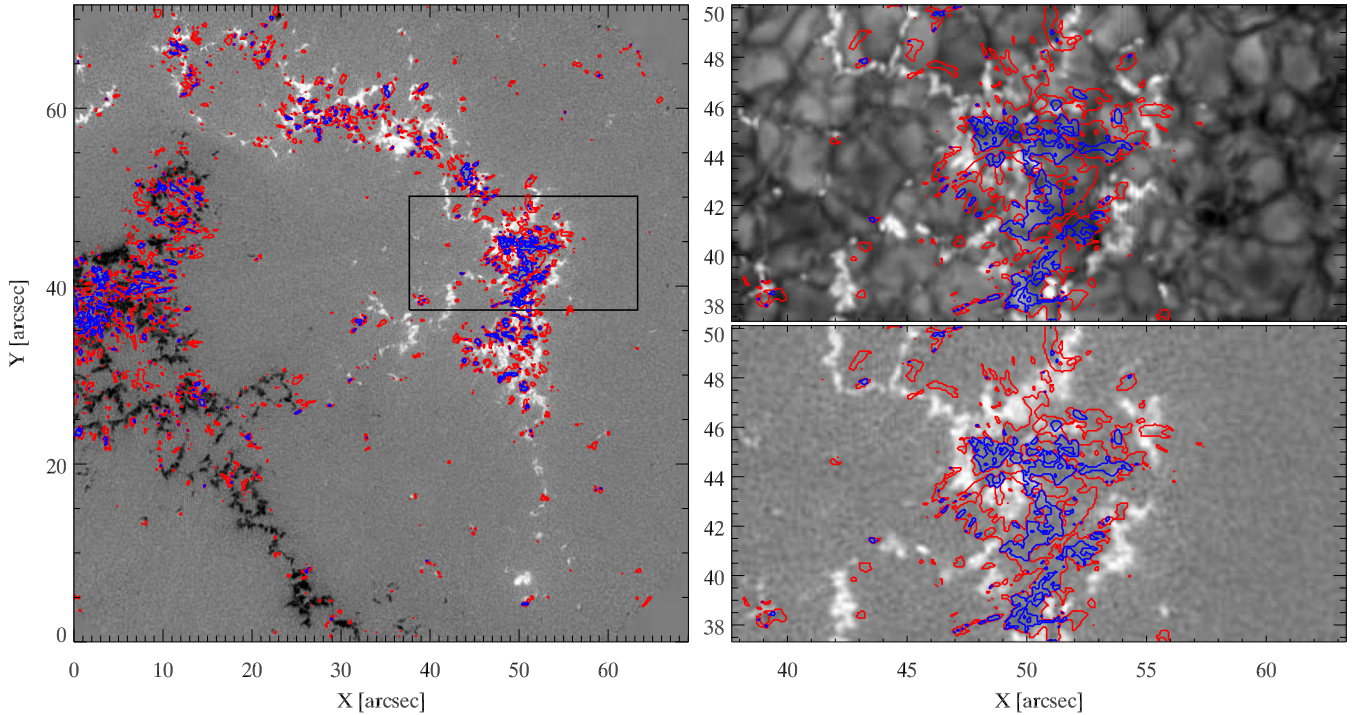}
\caption{Spatial distribution of RC profiles measured as number of zero-crossings of ${\rm d}I/{\rm d}\lambda$ over the core region ($\pm 582$~m\AA). The left panel shows a Stokes~$V$ map integrated over all 29 line positions with contours delineating areas with 3 ({\it red}) and 4 ({\it blue}) zero-crossings. The rightmost panels show a zoom-in on the region outlined in the left panel, with a wide-band image as background at the top. An online movie is available.}
\label{fig:f2}
\end{figure*}

The observed field-of-view (FOV) contains two patches of network with
opposite polarities. In the photosphere, observed in the wings of the
$\lambda8542$ line, the images show granulation with two
concentrations of bright-points. In the chromosphere, visible in the
core of the line, the images show a landscape of dark elongated fibrils
that extend outwards from the network patches. 

A section of our FOV is enlarged for clarity in
Fig.~\ref{fig:f1}. This patch of mostly unipolar plage shows strong magnetic field
concentrations (Stokes~$V$, top middle panel) that are tightly
correlated with the locations of bright-points (Stokes~$I$, top left panel). 
The line core image (top right panel), which is
sensitive to chromospheric plasma, shows fibrils extending from
the plage region. The plage region is generally very bright compared
with the fibrils.

We distinguish three different ``typical'' line profiles and give
examples from three different locations for each profile type: quiet
(locations and spectra marked with green
symbols), bright-point (blue symbols) and RC profiles (orange symbols).
The quiet profiles show a ``classical'' \ion{Ca}{2}~$\lambda8542$ profile
with bright photospheric wings and a chromospheric absorption core
(green profiles in bottom row of Fig.~\ref{fig:f1}). 
The shape of the bright-point profiles (blue in bottom row) is very similar to that of the
quiet-Sun profiles but with enhanced intensity for all wavelengths.

The RC profiles look markedly different. The photospheric wings of RC
profiles are virtually identical in brightness and shape to those of
quiet profiles. However, the chromospheric
core of RC profiles is very different: it is almost flat, and 
much brighter than the core of the quiet profiles. The
chromospheric core of RC profiles actually has an intensity that is similar to
that of the core of the bright-point profiles.

In a large fraction of the RC profiles, we have also noticed two weak
emission lobes at approximately
 $\lambda=\pm 194$~m\AA \ from line center. These lobes are further discussed in \S\S~\ref{sec:temporal} and \ref{sec:simulation}.

\subsection{Spatial distribution and ubiquity}\label{sec:ubi}

One clue to the cause of the RC profiles comes from investigating their
spatial distribution and ubiquity in the plage region. Clearly not all locations
show RC profiles. To determine, in an automated fashion, the location of RC profiles we tried several approaches. We found that the ratio between core and wing intensities reveals many locations of raised core profiles, but visual inspection shows that it also flags many "false positive" locations (usually narrow absorbing cores that are strongly Doppler-shifted). A better method exploits the fact that around the core of these
profiles the intensity is relatively flat with wavelength, so that the
derivative of intensity with wavelength $\text{d}I/\text{d}\lambda$ hovers around 0, with frequent
changes of the sign of the derivative. 
We thus developed an algorithm that
calculates the number of zero-crossings of $\text{d}I/\text{d}\lambda$
around the core of the line. Both bright-point and quiet profiles have
values of 1, whereas RC profiles almost always have larger values.

By overplotting contours of the number of
zero-crossings of $\text{d}I/\text{d}\lambda$  (red and blue contours
in Fig.~\ref{fig:f2}), we find that the RC profiles are located in the proximity of
photospheric bright-points and in extended patches of photospheric
granulation, which are embedded in patches of plage. In fact, they are
quite rare outside this configuration. This is quite clear from the
enlarged regions shown in the right column of Fig.~\ref{fig:f2}. The RC profiles are preferentially
located inside the ring of bright-points, with only a few exceptions
found outside. Our analysis also shows that these RC profiles are
ubiquitous and occur wherever there is plage (or network), with their
frequency increasing in more complex regions. The spatial extent of
seemingly coherent or contiguous regions with RC profiles is of order
a fraction of an arcsecond up to an arcsecond.

\subsection{Temporal properties}\label{sec:temporal}
\begin{figure*}
\centering
\includegraphics[width=0.85\hsize]{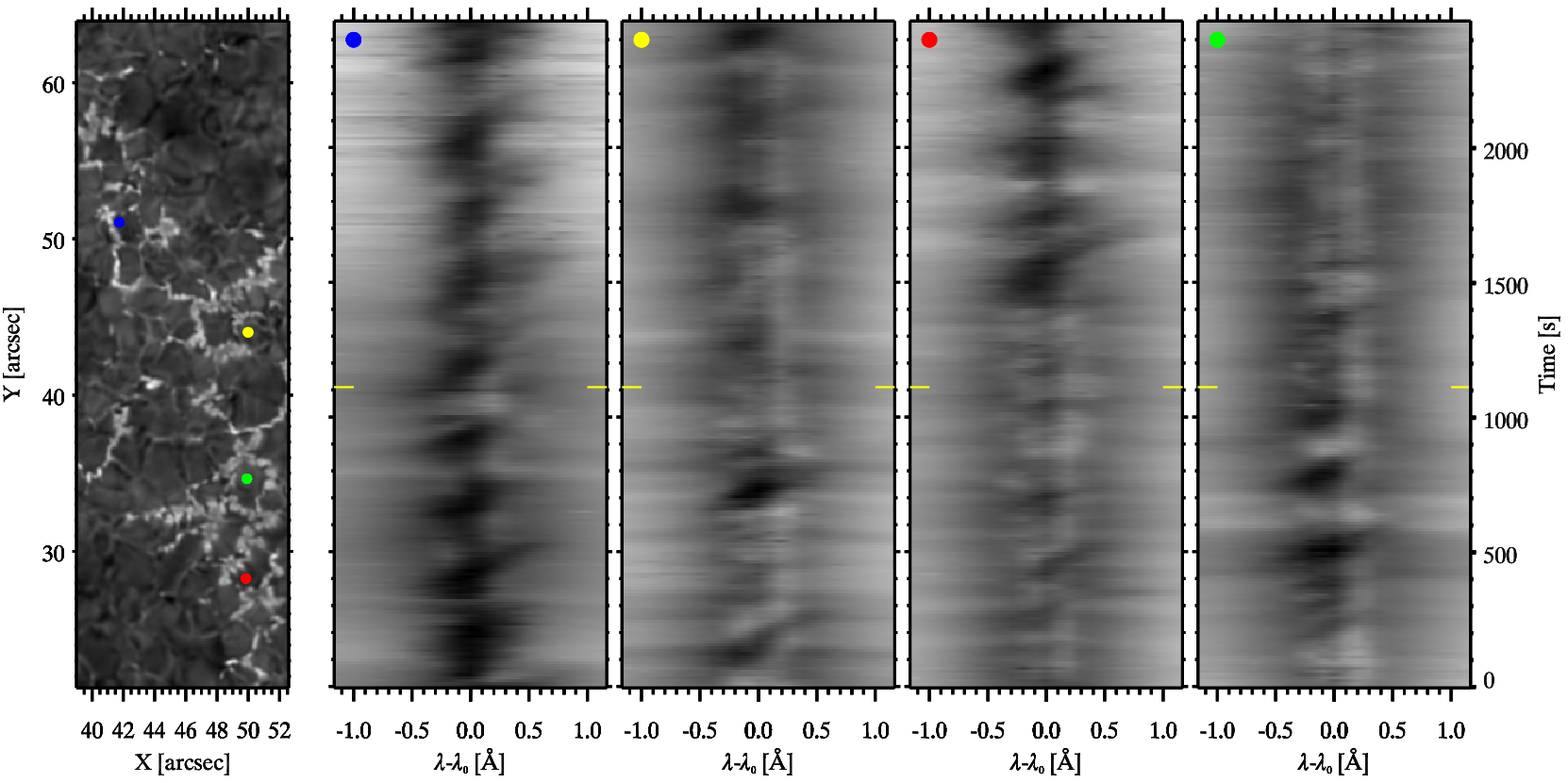}
\includegraphics[width=0.85\hsize]{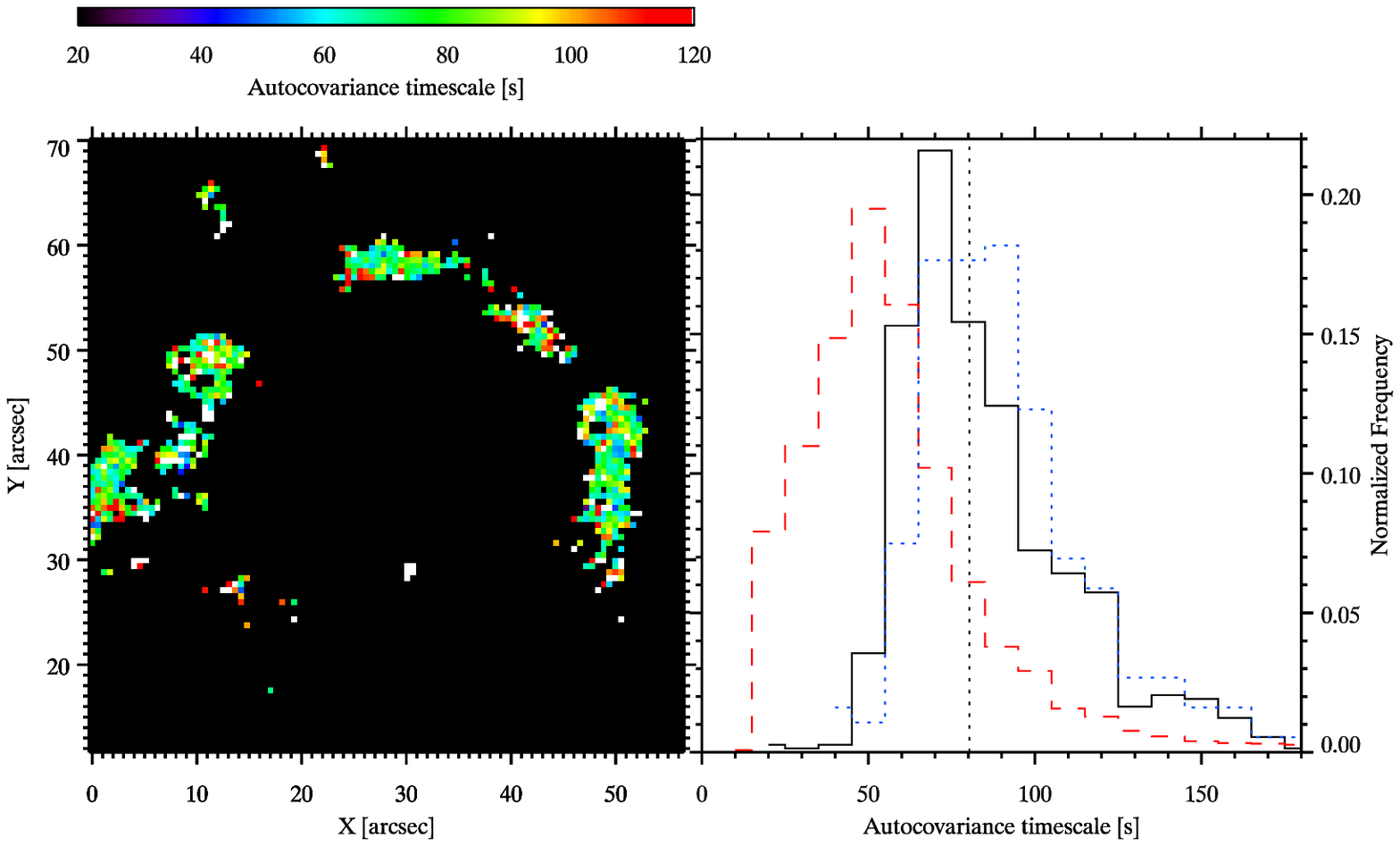}
\caption{Temporal evolution and lifetime of the RC profiles. The top
  row illustrates the temporal evolution of some $\lambda8542$
  profiles at locations indicated with color markers in the top left
  panel and repeated in each $\lambda-t$ plot (right panels). The second panel
from left corresponds to a traditional absorption profile showing a
shock induced temporal pattern. The last three panels correspond to
the time evolution of three RC profiles. The time of the image in the top-left panel is
indicated in the time slices with a yellow marker. 
The bottom row
provides information about the typical timescales that dominate the
temporal evolution of the RC
profiles:  map of the auto-covariance timescale ($8\times$ rebinned) ({\it left}) and a
histogram of auto-covariance timescales for locations with RC profiles ({\it right}), using
the original data ({\it dashed-red}),  
$8\times8$ ({\it solid-black}) and $16\times16$ rebinning ({\it dotted-blue}). The vertical
black-dotted line indicates the median auto-covariance timescale: 81 seconds.}
\label{fig:f3}
\end{figure*}

The temporal behavior of the RC profiles is very different from that
of a more typical profile, as can be seen by comparing the three
top-right panels with the second panel from the
left in the top row of Fig.~\ref{fig:f3}. The latter shows the
temporal evolution of the \ion{Ca}{2} $\lambda$8542 line profile in
a location that does not show strong RC behavior. It is dominated by a
succession of magneto-acoustic shocks, visible as strongly absorbing
features and similar to those described by
\citet{Langangen2008,2009vecchio}. Two brief moments
of RC profiles occur in between shocks. The three RC locations
(rightmost panels, top row) on the other hand
show raised core emission for much of the time-series, with only
occasional shock-induced absorption. The RC profiles are not constant in
time but show significant variability.
{ We find that the imprint from the red emission lobe at $\pm194$~m\AA \  is
 often noticeable throughout the timeseries, which is not always the
 case for the blue lobe.}
\begin{figure*}[t]
\includegraphics[width=\hsize]{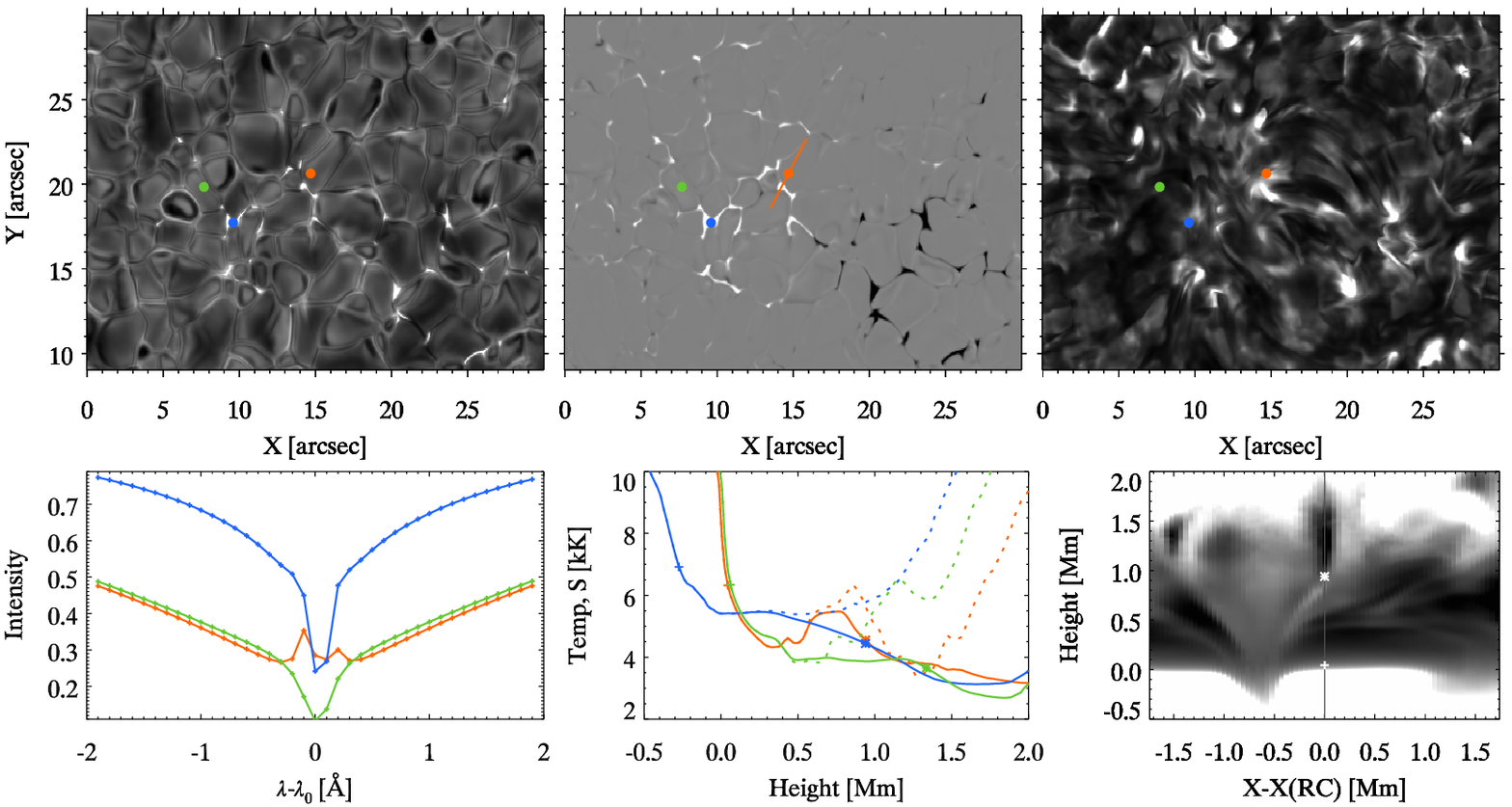}
\caption{Synthetic $\lambda8542$ observations and physical quantities from our 3D
  MHD simulation. In the top row, monochromatic intensity in the wing ({\it left}), the vertical
component of the magnetic field at $z=0$~km ({\it middle}) and the monochromatic
intensity at line center ({\it right}). The lower-left panel shows the intensity profiles at locations
indicated with colored markers in the top row: quiet-Sun ({\it green}), bright-point ({\it blue})
and raised-core profile ({\it orange}). The lower-middle panel
shows the temperature stratification ({\it dashed}) and source 
function ({\it solid}) at each location.
The markers over-plotted on the source function indicate the height at
which $\tau_\nu=1$ in the continuum ({\it plus}) and at line center ({\it star}). The lower right panel shows a 2D cut (along the orange line shown in the top middle panel) of the temperature as a
function of height in the simulation. The zero-point in x is at the
location of the orange circle in the top middle panel, the site of an
RC profile. Markers show the location of
$\tau_\nu=1$ in the continuum ({\it plus}) and at line center ({\it star}). }
\label{fig:f4}
\end{figure*}

The online movie of Fig.~\ref{fig:f2} shows evidence of
contiguous and/or coherent spatial regions that contain RC profiles
with typical spatial and temporal scales of order 0.5\arcsec and 1--2
minutes. These regions change shape and move about within the plage
region.
To determine the typical timescales over which RC profiles change,
we computed, for each RC location, the auto-covariance function of a
timeseries that is equal to 1 when the number of zero-crossings is
larger than 1, and otherwise 0. The auto-covariance timescale for that location is then 
the full width half max of this function. 

To make sure our timescale reflects the intrinsic lifetime and is not
dominated by either seeing-induced or proper motion of
the coherent RC region within the FOV, we calculated the auto-covariance timescale for data that was
rebinned with a range of different parameters (from $2\times2$ to $16\times16$
pixels). The timescales increase as the rebinning parameters
are increased from $1\times1$ to $16\times16$ (Fig.~\ref{fig:f3}, lower right panel).
This is because the timescales of the
data at the original resolution are dominated by seeing deformations,
jitter and proper motions leading to short lifetimes. 
Once we rebin the data $8\times8$ (solid-black)
we no longer see a difference in timescale compared to $16\times16$ (dotted-blue)
rebinned data, with both showing median values of about 80 seconds. 
This means that the intrinsic dominant lifetime of these coherent RC regions is about 1.5 minutes.
It also implies that the typical spatial scales are less than 8
pixels, i.e., $8\times0.07 \sim0.5$\arcsec.


\section{3D MHD Simulation} 
\label{sec:simulation}
We use a snapshot from a 3D MHD simulation calculated with
the \textsc{Bifrost} code \citep{2011gudiksen}. The simulation has 
an extent of
$24\times24\times16.8$~Mm, encompassing the upper convection zone,
photosphere, chromosphere and
corona. In the vertical direction, the simulation extends from 2.4~Mm
below to 14.4~Mm above average optical depth unity at
$\lambda=500$~nm. The grid spacing is 48~km horizontally and
non-equidistant vertically with a spacing of 19~km between
$z=-1$ and 5~Mm and increasing towards the upper and lower boundaries.
The simulation includes optically thick radiative transfer including scattering
in the photosphere and low chromosphere,
parametrized non-LTE radiative losses in the upper chromosphere, transition region
and corona, thermal conduction along magnetic field lines and an EOS that
includes the effects of non-equilibrium ionization of hydrogen, see
\citet{2011gudiksen} and  \citet{2012carlsson} for details. 
The magnetic field in the simulation consists of two patches of different
polarity separated by 8~Mm  (Fig.~\ref{fig:f4}, top middle panel). The field
is rather weak compared with the plage-like regions of the observations;
the average unsigned flux in the photosphere is 50~G.
The same snapshot was used in \citet{2012leenaarts}.

The synthetic $\lambda8542$ profiles are computed using the
code \textsc{Multi} \citep{Carlsson1986}, assuming each vertical column in the 3D snapshot to be an
independent 1D plane-parallel atmosphere. 
The profiles have been convolved with the CRISP spectral PSF 
(FWHM of 111~m\AA).

Our synthetic profiles show
evidence of quiet, bright-point and RC profiles (Fig.~\ref{fig:f4}) that are very similar to the observations. By
comparing the physical variables and radiative transfer properties for
various locations and profiles, we can investigate the physical cause for
the shape of the latter. 

At each wavelength, the intensity is approximately equal to the source function 
at monochromatic optical depth unity (the Eddington-Barbier relation). The
lower middle panel of Fig~\ref{fig:f4} shows the source function at the
three chosen locations with optical depth unity marked for the continuum and
for the line center. The intensity in the wings of the bright-point profile
is high because the strong magnetic flux regions are evacuated compared to
their surroundings to stay in horizontal total pressure balance. As a result,
we find lower densities and therefore a deeper formation { \citep{2001steiner,2003schussler,2004shelyag,2004carlsson}} where the temperature 
is higher (plus-signs in the lower-middle panel). Both the bright-point and 
quiet-Sun locations have monotonically decreasing source functions with 
height, giving pure absorption lines, because the source function decouples
from the Planck function lower in the atmosphere than the temperature increase. At the RC 
location we go from a quiet-Sun temperature to an increased temperature
typical of a magnetic region when we cross the expanding magnetic
field at about 0.5~Mm height (see also lower-right panel). The density is still high enough to ensure
a coupling between the Planck function and the source function and we get 
a local maximum of the latter, leading to two weak emission peaks in 
the RC profile at roughly $\pm200$~m\AA. { Therefore,
non-LTE effects and chromospheric heating can explain the presence of 
those emission peaks.}
The line core is formed above the local source function maximum
and there is therefore a central reversal. 
The optical depth unity at line
center is at equal heights for the bright-point location and the RC location
(because both locations are basically magnetic above the canopy height)
and the source functions are also almost the same, giving similar
line core intensity at the two locations. 

What makes RC profiles special is that they reveal the locations of a
sharp temperature gradient with height between a relatively quiet
photosphere and high temperature chromosphere. In the simulations this occurs
in the immediate vicinity of bright-points where we see the hot canopy
of magnetic field that expands with height from its roots in the
bright-points (lower-right panel, Fig.~\ref{fig:f4}). 

{ Our simulations also help explain why our detection method of
zero-crossings works well in identifying RC profiles, given that the two emission lobes
are naturally produced by a hot canopy overlying the quiet photosphere.}
\section{Discussion and conclusions}
\label{sec:discussion}

In this letter we characterize the observational properties of
profiles of the \ion{Ca}{2} $\lambda$8542 line in which the core emission is raised
compared to typical, quiet, profiles. These RC profiles have
photospheric wings similar to those of quiet profiles, but line-core
intensities that are as strong as in the profiles
from photospheric bright-points. The RC profiles are ubiquitous
throughout plage and network regions and predominantly located in the surroundings of
bright-points, usually on the inside of the plage region. We find that regions of
RC profiles show coherent spatial and temporal behavior on scales of
0.5\arcsec and 1.5 minutes, respectively.

We use RC profiles in synthetic observations computed from a
3D radiative MHD simulation to determine what causes the peculiar line
profile. The simulation shows that RC profiles occur for locations
with a steep increase in
the temperature stratification of about 1500~K, between $z\approx0.5$~Mm and
$z\approx1.0$~Mm. The reduced line opacity (due to the
vacuum effect of magnetic pressure) has a contributive effect to the
line core brightening.

In the simulations the strong temperature gradient occurs because 
the line of sight jumps from a quiet photosphere dominated
by granulation to the hot chromospheric canopy associated with flux
concentrations. These concentrations expand with height from their
photospheric bright-point roots to a volume filling field higher up.
Chromospheric heating is caused by strong
field gradients and currents (and associated dissipation) in
the vicinity of flux concentrations. This scenario is fully compatible
with our observations that show RC profiles in the immediate vicinity
(but not on top) of bright-points, and can explain the enhanced line core
intensity over the entire plage region. 

Our observations provide strict constraints on the location and
spatio-temporal properties of chromospheric heating in magnetic
regions. The observations fit well with the chromospheric heating present in our simulations, which 
is dominated by current dissipation. While the spatial scale of the
dissipation is mostly set by numerical resistivity in the simulation, 
recent work suggests that
ion-neutral interactions lead to a Pedersen resistivity
that has the same order of magnitude as the numerical resistivity,
thus rendering these simulations surprisingly close to solar
conditions \citep{Juan2012}. 
Our observations of apparently strong heating at the
interface of interacting flux tubes are also compatible with the
current-driven heating in the \textsc{Bifrost} models. Future work will have to clarify which of the
several other physical mechanisms proposed to drive heating in the magnetic
chromosphere is compatible with the observed properties, whether it is
high frequency wave heating \citep{Hasan2008, Vigeesh2009},
reconnection related to weak granular fields
\citep{Lites2008,Isobe2008}, or a turbulent cascade of Alfv\'en waves
\citep{vanBalle2011}.

Finally, we note that the rather flat line core
of the RC profiles means that the
weak-field approximation (where the field strength is inversely
proportional to $\text{d}I/\text{d}\lambda$) breaks down, leading to strong noise in the
derived magnetic field values.

\acknowledgments
The Swedish 1-m Solar Telescope is operated by the Institute for Solar
Physics of the Royal Swedish Academy of Sciences in the Spanish
Observatorio del Roque de los Muchachos of the Instituto de
Astrof\'{\i}sica de Canarias. 
The research has received funding from the Research Council
of Norway and
from the European Research
Council under the European Union's Seventh Framework Programme (FP7/2007-2013)/ERC 
Grant agreement n$^o$ 291058.
B.D.P. was supported through NASA grants NNX08BA99G, NNX08AH45G and
NNX11AN98G.  The authors gratefully acknowledge support from the
International Space Science Institute.


\begin{thebibliography}{39}
\expandafter\ifx\csname natexlab\endcsname\relax\def\natexlab#1{#1}\fi

\bibitem[{{Anderson} \& {Athay}(1989)}]{Anderson1989}
{Anderson}, L.~S., \& {Athay}, R.~G. 1989, \apj, 336, 1089

\bibitem[{Carlsson(1986)}]{Carlsson1986}
Carlsson, M. 1986, A Computer Program for Solving Multi-Level Non-LTE Radiative
  Transfer Problems in Moving or Static Atmospheres (Uppsala Astronomical
  Observatory: Report No.\ 33)

\bibitem[{{Carlsson} \& {Leenaarts}(2012)}]{2012carlsson}
{Carlsson}, M., \& {Leenaarts}, J. 2012, \aap, 539, A39

\bibitem[{{Carlsson} \& {Stein}(1997)}]{Carlsson1997}
{Carlsson}, M., \& {Stein}, R.~F. 1997, \apj, 481, 500

\bibitem[{{Carlsson} {et~al.}(2004){Carlsson}, {Stein}, {Nordlund}, \&
  {Scharmer}}]{2004carlsson}
{Carlsson}, M., {Stein}, R.~F., {Nordlund}, {\AA}., \& {Scharmer}, G.~B. 2004,
  \apjl, 610, L137

\bibitem[{{Cauzzi} {et~al.}(2009){Cauzzi}, {Reardon}, {Rutten}, {Tritschler},
  \& {Uitenbroek}}]{2009cauzzi}
{Cauzzi}, G., {Reardon}, K., {Rutten}, R.~J., {Tritschler}, A., \&
  {Uitenbroek}, H. 2009, \aap, 503, 577

\bibitem[{de~la Cruz~Rodr\'iguez(2010)}]{2010delacruz}
de~la Cruz~Rodr\'iguez, J. 2010, PhD thesis, Stockholm University, Department
  of Astronomy, http://urn.kb.se/resolve?urn=urn:nbn:se:su:diva-43646

\bibitem[{{De Pontieu} {et~al.}(2012){De Pontieu}, {Carlsson}, {Rouppe van der
  Voort}, {Rutten}, {Hansteen}, \& {Watanabe}}]{BDP2012}
{De Pontieu}, B., {Carlsson}, M., {Rouppe van der Voort}, L.~H.~M., {Rutten},
  R.~J., {Hansteen}, V.~H., \& {Watanabe}, H. 2012, \apjl, 752, L12

\bibitem[{{Fontenla} {et~al.}(1993){Fontenla}, {Avrett}, \&
  {Loeser}}]{Fontenla1993}
{Fontenla}, J.~M., {Avrett}, E.~H., \& {Loeser}, R. 1993, \apj, 406, 319

\bibitem[{{Goodman} \& {Kazeminezhad}(2010)}]{Goodman2010}
{Goodman}, M.~L., \& {Kazeminezhad}, F. 2010, \apj, 708, 268

\bibitem[{{Gudiksen} {et~al.}(2011){Gudiksen}, {Carlsson}, {Hansteen}, {Hayek},
  {Leenaarts}, \& {Mart{\'{\i}}nez-Sykora}}]{2011gudiksen}
{Gudiksen}, B.~V., {Carlsson}, M., {Hansteen}, V.~H., {Hayek}, W., {Leenaarts},
  J., \& {Mart{\'{\i}}nez-Sykora}, J. 2011, \aap, 531, A154

\bibitem[{{Hasan} \& {van Ballegooijen}(2008)}]{Hasan2008}
{Hasan}, S.~S., \& {van Ballegooijen}, A.~A. 2008, \apj, 680, 1542

\bibitem[{{Henriques}(2012)}]{2012henriques}
{Henriques}, V.~M.~J. 2012, \aap, 548, A114

\bibitem[{{Isobe} {et~al.}(2008){Isobe}, {Proctor}, \& {Weiss}}]{Isobe2008}
{Isobe}, H., {Proctor}, M.~R.~E., \& {Weiss}, N.~O. 2008, \apjl, 679, L57

\bibitem[{{Judge} {et~al.}(2010){Judge}, {Kn{\"o}lker}, {Schmidt}, \&
  {Steiner}}]{2010judge}
{Judge}, P., {Kn{\"o}lker}, M., {Schmidt}, W., \& {Steiner}, O. 2010, \apj,
  720, 776

\bibitem[{{Kleint}(2012)}]{2012kleint}
{Kleint}, L. 2012, \apj, 748, 138

\bibitem[{{Langangen} {et~al.}(2008){Langangen}, {Carlsson}, {Rouppe van der
  Voort}, {Hansteen}, \& {De Pontieu}}]{Langangen2008}
{Langangen}, {\O}., {Carlsson}, M., {Rouppe van der Voort}, L., {Hansteen}, V.,
  \& {De Pontieu}, B. 2008, \apj, 673, 1194

\bibitem[{{Leenaarts} {et~al.}(2012){Leenaarts}, {Carlsson}, \& {Rouppe van der
  Voort}}]{2012leenaarts}
{Leenaarts}, J., {Carlsson}, M., \& {Rouppe van der Voort}, L. 2012, \apj, 749,
  136

\bibitem[{{Lites} {et~al.}(2008){Lites}, {Kubo}, {Socas-Navarro}, {Berger},
  {Frank}, {Shine}, {Tarbell}, {Title}, {Ichimoto}, {Katsukawa}, {Tsuneta},
  {Suematsu}, {Shimizu}, \& {Nagata}}]{Lites2008}
{Lites}, B.~W., {et~al.} 2008, \apj, 672, 1237

\bibitem[{{L{\'o}pez Ariste} {et~al.}(2001){L{\'o}pez Ariste}, {Socas-Navarro},
  \& {Molodij}}]{2001lopez-ariste}
{L{\'o}pez Ariste}, A., {Socas-Navarro}, H., \& {Molodij}, G. 2001, \apj, 552,
  871

\bibitem[{{Mart{\'{\i}}nez-Sykora} {et~al.}(2012){Mart{\'{\i}}nez-Sykora}, {De
  Pontieu}, \& {Hansteen}}]{Juan2012}
{Mart{\'{\i}}nez-Sykora}, J., {De Pontieu}, B., \& {Hansteen}, V. 2012, \apj,
  753, 161

\bibitem[{{Pietarila} {et~al.}(2007{\natexlab{a}}){Pietarila}, {Socas-Navarro},
  \& {Bogdan}}]{2007pietarilab}
{Pietarila}, A., {Socas-Navarro}, H., \& {Bogdan}, T. 2007{\natexlab{a}}, \apj,
  670, 885

\bibitem[{{Pietarila} {et~al.}(2007{\natexlab{b}}){Pietarila}, {Socas-Navarro},
  \& {Bogdan}}]{2007pietarilaa}
---. 2007{\natexlab{b}}, \apj, 663, 1386

\bibitem[{{Rouppe van der Voort} {et~al.}(2009){Rouppe van der Voort},
  {Leenaarts}, {de Pontieu}, {Carlsson}, \& {Vissers}}]{Rouppe2009}
{Rouppe van der Voort}, L., {Leenaarts}, J., {de Pontieu}, B., {Carlsson}, M.,
  \& {Vissers}, G. 2009, \apj, 705, 272

\bibitem[{{Scharmer} {et~al.}(2008){Scharmer}, {Narayan}, {Hillberg}, {de la
  Cruz Rodr{\'{\i}}guez}, {L{\"o}fdahl}, {Kiselman}, {S{\"u}tterlin}, {van
  Noort}, \& {Lagg}}]{2008ApJ...689L..69S}
{Scharmer}, G.~B., {et~al.} 2008, \apjl, 689, L69

\bibitem[{{Schnerr} {et~al.}(2011){Schnerr}, {de La Cruz Rodr{\'{\i}}guez}, \&
  {van Noort}}]{2011schnerr}
{Schnerr}, R.~S., {de La Cruz Rodr{\'{\i}}guez}, J., \& {van Noort}, M. 2011,
  \aap, 534, A45

\bibitem[{{Sch{\"u}ssler} {et~al.}(2003){Sch{\"u}ssler}, {Shelyag},
  {Berdyugina}, {V{\"o}gler}, \& {Solanki}}]{2003schussler}
{Sch{\"u}ssler}, M., {Shelyag}, S., {Berdyugina}, S., {V{\"o}gler}, A., \&
  {Solanki}, S.~K. 2003, \apjl, 597, L173

\bibitem[{{Shelyag} {et~al.}(2004){Shelyag}, {Sch{\"u}ssler}, {Solanki},
  {Berdyugina}, \& {V{\"o}gler}}]{2004shelyag}
{Shelyag}, S., {Sch{\"u}ssler}, M., {Solanki}, S.~K., {Berdyugina}, S.~V., \&
  {V{\"o}gler}, A. 2004, \aap, 427, 335

\bibitem[{{Socas-Navarro} {et~al.}(2000){Socas-Navarro}, {Trujillo Bueno}, \&
  {Ruiz Cobo}}]{2000socas-navarro-sci}
{Socas-Navarro}, H., {Trujillo Bueno}, J., \& {Ruiz Cobo}, B. 2000, Science,
  288, 1396

\bibitem[{{Steiner} {et~al.}(2001){Steiner}, {Hauschildt}, \&
  {Bruls}}]{2001steiner}
{Steiner}, O., {Hauschildt}, P.~H., \& {Bruls}, J. 2001, \aap, 372, L13

\bibitem[{{Straus} {et~al.}(2008){Straus}, {Fleck}, {Jefferies}, {Cauzzi},
  {McIntosh}, {Reardon}, {Severino}, \& {Steffen}}]{Straus2008}
{Straus}, T., {Fleck}, B., {Jefferies}, S.~M., {Cauzzi}, G., {McIntosh}, S.~W.,
  {Reardon}, K., {Severino}, G., \& {Steffen}, M. 2008, \apjl, 681, L125

\bibitem[{{van Ballegooijen} {et~al.}(2011){van Ballegooijen}, {Asgari-Targhi},
  {Cranmer}, \& {DeLuca}}]{vanBalle2011}
{van Ballegooijen}, A.~A., {Asgari-Targhi}, M., {Cranmer}, S.~R., \& {DeLuca},
  E.~E. 2011, \apj, 736, 3

\bibitem[{{van Noort} {et~al.}(2005){van Noort}, {Rouppe van der Voort}, \&
  {L{\"o}fdahl}}]{2005SoPh..228..191V}
{van Noort}, M., {Rouppe van der Voort}, L., \& {L{\"o}fdahl}, M.~G. 2005,
  \solphys, 228, 191

\bibitem[{{Vecchio} {et~al.}(2009){Vecchio}, {Cauzzi}, \&
  {Reardon}}]{2009vecchio}
{Vecchio}, A., {Cauzzi}, G., \& {Reardon}, K.~P. 2009, \aap, 494, 269

\bibitem[{{Vernazza} {et~al.}(1981){Vernazza}, {Avrett}, \&
  {Loeser}}]{Vernazza1981}
{Vernazza}, J.~E., {Avrett}, E.~H., \& {Loeser}, R. 1981, \apjs, 45, 635

\bibitem[{{Vigeesh} {et~al.}(2009){Vigeesh}, {Hasan}, \&
  {Steiner}}]{Vigeesh2009}
{Vigeesh}, G., {Hasan}, S.~S., \& {Steiner}, O. 2009, \aap, 508, 951

\bibitem[{{Vissers} \& {Rouppe van der Voort}(2012)}]{2012vissers}
{Vissers}, G., \& {Rouppe van der Voort}, L. 2012, \apj, 750, 22

\bibitem[{{Wedemeyer-B{\"o}hm} {et~al.}(2012){Wedemeyer-B{\"o}hm}, {Scullion},
  {Steiner}, {Rouppe van der Voort}, {de La Cruz Rodriguez}, {Fedun}, \&
  {Erd{\'e}lyi}}]{2012wedemeyer}
{Wedemeyer-B{\"o}hm}, S., {Scullion}, E., {Steiner}, O., {Rouppe van der
  Voort}, L., {de La Cruz Rodriguez}, J., {Fedun}, V., \& {Erd{\'e}lyi}, R.
  2012, \nat, 486, 505

\bibitem[{{Withbroe} \& {Noyes}(1977)}]{Withbroe1977}
{Withbroe}, G.~L., \& {Noyes}, R.~W. 1977, \araa, 15, 363

\end{thebibliography}

\end{document}